\documentclass[draft,preprint,aps,%superscriptaddress,
showpacs%,nofootinbib,eqsecnum,preprintnumbers,twocolumn,prl
]{revtex4}

\usepackage{amsmath}
\usepackage{amssymb}
\usepackage{graphicx}
\usepackage{bm}
\usepackage{epsfig}
\input{epsf}
\begin{document}

\preprint{Guchi-TP-021}

\title{
One-loop effective potential for the vacuum gauge field in
$M_3\times S^3 \times S^1$ space-time
}
\author{Yoshinori Cho}
\email{b2669@sty.cc.yamaguchi-u.ac.jp}
\affiliation{
Graduate School of Science and Engineering, Yamaguchi University
\\
Yoshida, Yamaguchi-shi, Yamaguchi 753-8512, Japan
}
\author{Kiyoshi Shiraishi}
\email{shiraish@sci.yamaguchi-u.ac.jp}
\affiliation{
Faculty of Science, Yamaguchi University
\\ Yoshida, Yamaguchi-shi, Yamaguchi 753-8512, Japan
}

\date{\today}
\begin{abstract}
%%%%%%%%%%%%%%%%%%%%%%%%%%%%%%%%%%%%%%%%%%%%%%%%%%%%%%%%%%%%%%%%%%%%%
We calculate the effective potential
for the vacuum gauge field from the one-loop matter effect
in $M_3\times S^3\times S^1$ space-time.
This background geometry is motivated from the recent
studies on gauged supergravities with a positive-definite
potential, which admits a generalized Kaluza-Klein reduction.
We investigate how symmetry breaking patterns through the Hosotani
mechanism are affected by the ratio of the radii of $S^1$ and $S^3$.
%%%%%%%%%%%%%%%%%%%%%%%%%%%%%%%%%%%%%%%%%%%%%%%%%%%%%%%%%%%%%%%%%%%%%
\end{abstract}

\maketitle
%%%%%%%%%%%%%%%%%%%%%%%%%%%%%%%%%%%%%%%%%%%%%%%%%%%%%%%%%%%%%%%%%%%%%
\section{Introduction}
%%%%%%%%%%%%%%%%%%%%%%%%%%%%%%%%%%%%%%%%%%%%%%%%%%%%%%%%%%%%%%%%%%%%%
Recently there has been much interest in a dynamical gauge
symmetry breaking mechanism in higher dimensional theories,
{\it i.e.}, the Hosotani mechanism considered on various compact
extra dimensions~\cite{hosotani}.
In this mechanism a constant vacuum gauge field on a
multiply-connected space or an orbifold,
which is related with the Wilson line element on the space,
becomes dynamical degrees of freedom
and develops the vacuum expectation value, which plays
a role of an ``order parameter" for gauge symmetry breaking.
Moreover the analysis of the quantum effects
in terms of the one-loop effective potential is crucial
for determining the order parameter.

For the non-local quantity like the Wilson line element,
the effective potential for the vacuum gauge field
does not depend on the ultraviolet effects
so that the massive particles seem to
affect the gauge symmetry breaking
patterns through the Hosotani mechanism.
In fact, Takenaga pointed out that the existence of bare mass terms
changes the gauge symmetry breaking patterns \cite{takenaga}.

In this paper, we consider how the symmetry patterns through
Hosotani mechanism are affected by the ratio of the radii of
$S^1$ and $S^3$ in $M_3\times S^3\times S^1$ space-time.

This background geometry is motivated from the recent studies
on gauged supergravities with a positive-definite potential
\cite{salam,cvetic,kerimo1,kerimo2,kerimo3}.
These studies were followed
by a generalized Kaluza-Klein reduction~\cite{pope},
leading to a positive-definite  potential.
Moreover the generalized Kaluza-Klein procedure is studied
in arbitrary dimensions in~\cite{kerimo3}.
In this method, the generalized Kaluza-Klein reduction
was carried out on the bosonic content of the half-maximal
supergravity in $D=d+1(\leq 10)$ which consists of the graviton,
the antisymmetric tensor, the vector field and the dilaton,
which appear in the bosonic parts of the heterotic string theory
(or the NS-NS sector of the type-II string theory).
The various $d$ dimensional supergravities resulted
from the generalized Kaluza-Klein reduction have
spontaneously compactified solutions to
$M_{d-3}\times S^3$ and $M_{d-2}\times S^2$ vacuum solutions,
assuming that the antisymmetric tensor field
has a monopole configuration on $S^3$ and $S^2$,
which guarantees that the rest of the space-time is flat.

In this paper, we consider a toy model for the Hosotani mechanism
in the space-time motivated by the seven-dimensional
gauged supergravity with a positive-definite potential,
admitting a  spontaneous compactification to
a $M_3 \times S^3 \times S^1$ vacuum solution.
We assume that the antisymmetric tensor field on $S^3$
forms the magnetic flux configuration,
while there is a gauge field on $S^1$ which becomes dynamical
degrees of freedom,
causing a symmetry breaking under a certain condition that
appeared in the Hosotani mechanism.
We further focus on our attention to
the possibility of the symmetry breaking
without specifying a group representation.

In such a background space-time, we consider quantum effects
of a scalar field as well as a fermion without
a coupling to the flux on $S^3$.
In this setting, we calculate the one-loop effective
potential for the vacuum gauge field
in $M_3\times S^3\times S^1$ space-time.
In the present model, there are no massless fermion
in the low-energy effective theory.
Thus the effects of the non-zero
eigenmodes of $S^3$ may be crucial for
the symmetry breaking patterns.

The organization of this paper is as follows;
in the next section, in order to estimate the effective potential
for the vacuum gauge field on $S^1$,
we compute the one-loop effects of matter fields
in $M_3\times S^3\times S^1$ space-time.
We investigate how the symmetry breaking patterns through the
Hosotani mechanism are affected by
changing the ratio of radii of $S^1$ and $S^3$,
after carried out the regularization
and the summation over the massive modes.
Finally we discuss the result in the last section.
%%%%%%%%%%%%%%%%%%%%%%%%%%%%%%%%%%%%%%%%%%%%%%%%%%%%%%%%%%%%%%%%%%%%%
%%%%%%%%%%%%%%%%%%%%%%%%%%%%%%%%%%%%%%%%%%%%%%%%%%%%%%%%%%%%%%%%%%%%%
\section{One-loop effective potential}
%%%%%%%%%%%%%%%%%%%%%%%%%%%%%%%%%%%%%%%%%%%%%%%%%%%%%%%%%%%%%%%%%%%%%
In this section, we consider a toy model motivated by
the seven-dimensional gauged supergravity with
a positive-definite potential, permitting a vacuum
solution whose space-time geometry
is $M_3\times S^3 \times S^1$.
We compute the quantum effects of the scalar and fermion field with
the vacuum gauge field in this background solution.
We restrict our attention to whether the symmetry is broken
or not, without concerning about the  symmetries realized.
%%%%%%%%%%%%%%%%%%%%%%%%%%%%%%%%%%%%%%%%%%%%%%%%%%%%%%%%%%%%%%%%%%%%%
%%%%%%%%%%%%%%%%%%%%%%%%%%%%%%%%%%%%%%%%%%%%%%%%%%%%%%%%%%%%%%%%%%%%%
\subsection{Scalar field}
%%%%%%%%%%%%%%%%%%%%%%%%%%%%%%%%%%%%%%%%%%%%%%%%%%%%%%%%%%%%%%%%%%%%%
Firstly, we compute the one-loop quantum effect of
the scalar field in $M_3\times S^3 \times S^1$ space-time.
We assume that the scalar field is decoupled from
the flux on $S^3$.
In general, as is denoted in~\cite{ref13},
the effective vacuum energy density $V_{0}$ in
a background geometry which is $M_{(d+1)}\times S^N$ space-time
[($d+1$)-dimensional Minkowski space-time $\times$ $N$-sphere]
is represented by
\begin{eqnarray}
 V_{0}& \propto &
 \frac{1}{2}\ln \det [p^2+m^2] \nonumber \\
 &=&
 \frac{1}{2}\int\frac{d^{d+1}p}{(2\pi)^{d+1}}\sum^{\infty}_{l=0}
 D^{\rm sca}_l(N)\ln
 \left(
 p^2-\frac{\Lambda^{\rm sca}_l(N)}{a^2}+m^2
 \right), \\
 &=&
 -\frac{1}{2}\int^{\infty}_0
 \frac{dt}{(4\pi)^{\frac{d+1}{2}}}
 \frac{1}{t^{\frac{d+1}{2}+1}}\sum^{\infty}_{l=0}
 D^{\rm sca}_l(N)
 \exp\left[
 \left(
 \frac{\Lambda^{\rm sca}_l(N)}{a^2}-m^2
 \right)t
 \right],
\end{eqnarray}
for a single scalar degree of freedom.
Here $a$ is the radius of $S^N$ and
$m$ is a bare mass. $D^{\rm sca}_l(N)$ and $-\Lambda^{\rm sca}_l(N)$
denote the degeneracies and the eigenvalues of the laplacian
on the unit $N$-sphere, respectively.

For a scalar field, the eigenvalues of the laplacian and
the degeneracies on $S^N$ are obtained by
\begin{equation}
 \Lambda^{\rm sca}_l(n)
 =-l(l+n+1), \qquad
 D^{\rm sca}_l(n)=
 \frac{(2l+n+1)\Gamma(l+n-1)}{l!\Gamma(n)},
\end{equation}
respectively.

In addition, if we consider the $S^1$ compactification,
we take the additional discrete eigenvalues $n^2/b^2$,
where $n$ is integer and $b$ is the radius of $S^1$.
Then the one-loop vacuum energy in $M_3\times S^3 \times S^1$
space-time can be written as
\begin{equation}
 -\frac{1}{2(4\pi)^{\frac{3}{2}}}
 \int^{\infty}_0
 \frac{dt}{t^{\frac{5}{2}}}
 \sum^{\infty}_{l=0}
 \sum^{\infty}_{n=-\infty}
 D^{\rm sca}_l(3)
 \exp\left[
 \left(\frac{\Lambda^{\rm sca}_l(3)}{a^2}
 -m^2\right)t\right]
 \exp\left[-\frac{n^2}{b^2}t\right].
\end{equation}

Now we consider the vacuum gauge field on $S^1$.
Once we choose a symmetry group and a group representation
of the scalar field, the coupling to the vacuum gauge field
is determined and the eigenvalues on $S^1$ is specified for
each component field.
However, if we concentrate ourselves on the {\it possibility}
of the symmetry breaking, we need not specify the representation.
We assume that the spectrum of the scalar component coupled
to the non-zero vacuum gauge field should take the following form
\begin{equation}
 \frac{(n-v)^2}{b^2}\, ,
 \label{spectrum}
\end{equation}
where $v$ denotes a characteristic scale of the vacuum gauge field.
Thus the {\it necessary} condition for symmetry breaking is
the one-loop effective potential $V_{\rm eff}^{\rm sca}$, which
is obtained by
\begin{equation}
 V^{\rm sca}_{\rm eff}=
 -\frac{1}{2(4\pi)^{\frac{3}{2}}}\int^{\infty}_0
 \frac{dt}{t^{\frac{5}{2}}}\sum^{\infty}_{l=0}
 \sum^{\infty}_{n=-\infty}
 D^{\rm sca}_l(3)
 \exp\left[
 \left(\frac{\Lambda^{\rm sca}_l(3)}{a^2}
 -m^2\right)t\right]
 \exp\left[-\frac{(n-v)^2}{b^2}t\right],
\end{equation}
has the non-trivial minimum at non-zero $v$.
Note that this is not a {\it sufficient} condition.
Non-zero $v$ may be equivalent to the trivial vacuum,
which depends on the group representation~\cite{hosotani}.

We use the formula for the elliptic theta function:
\begin{equation}
 \sum^{\infty}_{n=-\infty}
 \exp\left[-\frac{a^2}{b^2}(n-v)^2 t\right]
 =\frac{b}{a}\sqrt{\frac{\pi}{t}}
 \sum^{\infty}_{n=-\infty} 
 \exp\left[-\frac{\pi^2 b^2n^2}{a^2 t}\right]
 \cos(2\pi nv).
 \label{theta function}
\end{equation}

Then one can find the one-loop effective potential as
\begin{eqnarray}
 V_{\rm eff}^{\rm sca}&=&-\frac{b}{16\pi a^4}\int^{\infty}_0
 \frac{dt}{t^3}e^{-(m^2a^2-1)t}\sqrt{\frac{\pi}{t}}
 \left[\frac{1}{4t} \right. \nonumber \\
 & & \left. +\sum^{\infty}_{l=0}
 \left(\frac{1}{2t}-\frac{\pi^2l^2}{t^2}\right)
 e^{-\frac{\pi^2l^2}{t}}\right]
 \sum^{\infty}_{n=-\infty}\exp\left[
 -\frac{\pi^2 b^2 n^2}{a^2 t}\right]
 \cos(2\pi n v).
\end{eqnarray}

In addition, we use an integral representation of the modified
Bessel function:
%in order to 
%obtain the expression in the condition of the 
%zero limit of the bare mass:
\begin{equation}
 K_{\nu}(z)=
 \frac{1}{2}\left(\frac{z}{2}\right)^{\nu}\int^{\infty}_0 
 \exp\left(-t-\frac{z^2}{4t}\right)t^{-\nu-1}dt.
 \label{bessel}
\end{equation}

Then one can rewrite the effective potential as follows:
\begin{eqnarray}
 V_{\rm eff}^{\rm sca}&=&
 -\frac{b}{16\sqrt{\pi}a^4}
 \sum^{\infty}_{l=0}\sum^{\infty}_{n=-\infty}
 \cos(2\pi nv)
 \left[
 \frac{1}{2}
 \left(
 \frac{(m^2a^2-1)}{\pi^2(b^2/a^2)n^2}
 \right)^{\frac{7}{4}}
 K_{\frac{7}{2}}\left(2\sqrt{(m^2a^2-1)
 (\pi^2(b^2/a^2)n^2)}\right)
 \nonumber \right. \\
 & &
 \left. +
 \left(\frac{(m^2a^2-1)}{\pi^2l^2+\pi^2(b^2/a^2)n^2}
 \right)^{\frac{7}{4}}K_{\frac{7}{2}}
 \left(2\sqrt{(m^2a^2-1)
 (\pi^2l^2+\pi^2(b^2/a^2)n^2)}\right)
 \nonumber \right. \\
 & &
 \left. 
 -2\pi^2l^2\left(\frac{(m^2a^2-1)}{\pi^2l^2+\pi^2(b^2/a^2)n^2}
 \right)^{\frac{9}{4}}K_{\frac{9}{2}}
 \left(2\sqrt{(m^2a^2-1)
 (\pi^2l^2+\pi^2(b^2/a^2)n^2)}\right)
 \right].
\end{eqnarray}

We consider the case of the massless scalar field and
drop the divergence which is independent of $v$.
Finally, we show the effective potential for
the various values of the radius of $S^3$ in FIG.~1.
The shape of the curve depends only on the ratio $b/a$.
The potential minimum is always located at $v=0$.
%%%%%%%%%%%%%%%%%%%%%%%%%%%%%%%%%%%%%%%%%%%%%%%%%%%%%%%%%%%%%%%%%%%%%
%%%%%%%%%%%%%%%%%%%%%%%%%%%%%%%%%%%%%%%%%%%%%%%%%%%%%%%%%%%%%%%%%%%%%
\subsection{Fermion field}
Next we perform the calculation of the one-loop quantum effect
of the fermion field in $M_3\times S^3 \times S^1$ space-time.
The methods of the calculation are similar to the case of
the scalar field, except for the eigenvalues of the laplacian
and the degeneracies on $S^N$ which are obtained by
\begin{equation}
 \Lambda_l^{\rm fer}(N)
 =\left(l+\frac{N}{2}\right)^2, \qquad
 D^{\rm fer}_l(N)=
 2^{\frac{N+1}{2}}\frac{\Gamma(l+N)}{l!\Gamma(N)},
\end{equation}
respectively.

As was discussed in the previous subsection,
the contribution of discrete eigenvalues of
$S^1$ and the vacuum gauge field
on $S^1$ can be taken by considering (\ref{spectrum}).

Then one can find the one-loop effective potential,
setting the bare mass to zero, as
\begin{eqnarray}
 V_{\rm eff}^{\rm fer}
 &=& 
 \frac{b}{(4\pi)^{\frac{1}{2}}a^4}\int^{\infty}_0 
 \frac{dt}{t^{\frac{7}{2}}}
 \left[\frac{1}{4t}-\frac{1}{8}
 \right.  \nonumber \\
 & &
 + \left. \sum^{\infty}_{l=0}(-1)^l 
 \left(
 -\frac{1}{4}+\frac{1}{2t}-\frac{\pi^2l^2}{t^2}
 \right)
 e^{-\frac{\pi^2l^2}{t}}
 \right]
 \sum^{\infty}_{n=-\infty}
 \exp\left[-\frac{\pi^2b^2n^2}{a^2t}
 \right]
 \cos(2\pi nv),
\end{eqnarray}
after using the formula for the elliptic theta
function~(\ref{theta function}).
This is normalized as a contribution of a single Dirac fermion.

We also drop the divergence independent of $v$.
Finally, we show the effective potential for the
various values of the  radius of $S^3$ in FIG.~2.
For sufficiently small $a/b$, the potential minimum appears at $v=0$.
%%%%%%%%%%%%%%%%%%%%%%%%%%%%%%%%%%%%%%%%%%%%%%%%%%%%%%%%%%%%%%%%%%%%%
%%%%%%%%%%%%%%%%%%%%%%%%%%%%%%%%%%%%%%%%%%%%%%%%%%%%%%%%%%%%%%%%%%%%%
\section{Summary}
%%%%%%%%%%%%%%%%%%%%%%%%%%%%%%%%%%%%%%%%%%%%%%%%%%%%%%%%%%%%%%%%%%%%%
In this paper, we have calculated the quantum effects
of the scalar and fermion field for the vacuum gauge field
in $M_3\times S^3 \times S^1$ space-time.
This background geometry can be obtained from
the spontaneous compactification of
the seven-dimensional gauged supergravity with a
positive-definite potential such as
the models constructed from the
generalized Kaluza-Klein reduction.

We have shown the one-loop effective potential
of the scalar field for various values of the radius
of $S^3$ in FIG.~1.
We can find that the symmetry breaking itself cannot appear,
for any values of the radius,
because the effective potential does not have
non-trivial minimum at non-zero $v$.
The reason is that there
is always the contribution of the zero-mode of $S^3$
to the effective potential.

On the contrary, for the one-loop effective potential
of the fermion field showed in FIG.~2,
one can find that there is the possibility of causing
the symmetry breaking in some cases, for example,
with the fermion in the adjoint representation.
The symmetry breaking patterns through the Hosotani mechanism,
therefore, change due to the effects of the massive modes of $S^3$.
This is because there is no zero-mode
in the spectrum of the effective theory
for the fermion field without the flux coupling.

Although we have considered the toy model started from
the motivation about the gauged supergravity,
the quantum effects of the effective theory are important
for the dynamical symmetry breaking
during the cosmological evolution.
Further, the supersymmetry breaking effect,
leading to the vacuum energy contribution
to the cosmological constant, must be taken into consideration.

In the present paper, we have treated only the fermion
without the flux coupling, and
then there is not a zero-mode in the spectrum.
We will further investigate the fermion with a flux coupling
having a zero-mode.
We anticipate that the symmetry breaking patterns
may not change as in the case with the scalar field.
%%%%%%%%%%%%%%%%%%%%%%%%%%%%%%%%%%%%%%%%%%%%%%%%%%%%%%%%%%%%%%%%%%%%%
%%%%%%%%%%%%%%%%%%%%%%%%%%%%%%%%%%%%%%%%%%%%%%%%%%%%%%%%%%%%%%%%%%%%%
\begin{figure}[h]
\epsfxsize=8cm
\mbox{\epsfbox{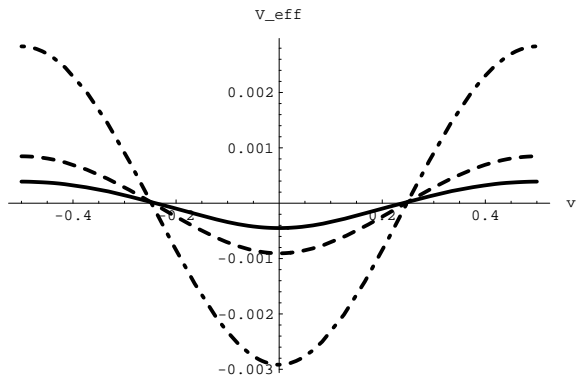}}
\bigskip
\caption{The one-loop effective potential $V^{\rm sca}_{\rm eff}$
of the scalar field is plotted against the
effects $v$ of the vacuum gauge field on $S^1$.
We set the radius of $S^1$ to unity ($b=1$).
The solid line corresponds to $a=1$,
the dashed line corresponds to $a=3$,
the dot-dashed line corresponds to $a=5$.
}
\end{figure}
\begin{figure}[h]
\epsfxsize=8cm
\mbox{\epsfbox{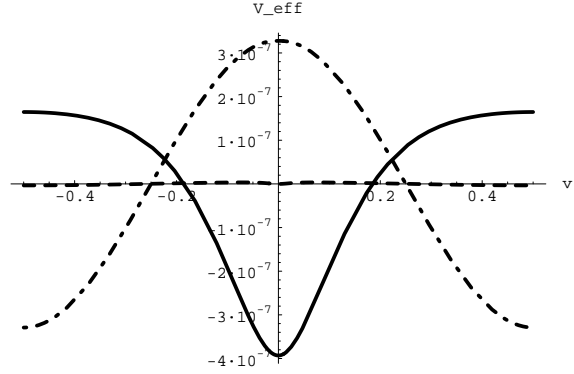}}
\caption{The one-loop effective potential $V^{\rm fer}_{\rm eff}$
of the fermion field is plotted against
the effects $v$ of the vacuum gauge field on $S^1$.
We set the radius of $S^1$ to unity ($b=1$).
The solid line corresponds to $a=0.1$,
the dashed line corresponds to $a=0.5$,
the dot-dashed line corresponds to $a=0.7$.
}
\end{figure}
%%%%%%%%%%%%%%%%%%%%%%%%%%%%%%%%%%%%%%%%%%%%%%%%%%%%%%%%%%%%%%%%%%%%%
%%%                    acknowledgments
%%%%%%%%%%%%%%%%%%%%%%%%%%%%%%%%%%%%%%%%%%%%%%%%%%%%%%%%%%%%%%%%%%%%%
\begin{acknowledgments}
The authors would like to thank N.~Kan and
R. Takakura for useful advice.
\end{acknowledgments}
%%%%%%%%%%%%%%%%%%%%%%%%%%%%%%%%%%%%%%%%%%%%%%%%%%%%%%%%%%%%%%%%%%%%%
%%%References
%%%%%%%%%%%%%%%%%%%%%%%%%%%%%%%%%%%%%%%%%%%%%%%%%%%%%%%%%%%%%%%%%%%%%

%%%%%%%%%%%%%%%%%%%%%%%%%%%%%%%%%%%%%%%%%%%%%%%%%%%%%%%%%%%%%%%%%%%%%
%%%%%%%%%%%%%%%%%%%%%%%%%%%%%%%%%%%%%%%%%%%%%%%%%%%%%%%%%%%%%%%%%%%%%
\end{document}